\documentclass[twocolumn,superscriptaddress,amsmath,amssymb,showpacs]{revtex4-1}
\usepackage[utf8x]{inputenc}

\usepackage{amsmath}
\usepackage{graphicx}
\usepackage{ulem}
\makeatletter
\usepackage{dcolumn}

\usepackage{physics}
\usepackage{color}

\makeatother

\begin{document}

\title{Maximum Quantum Entropy Method}

\begin{abstract}
Maximum entropy method for analytic continuation is extended by
introducing quantum relative entropy.  This new method is formulated
in terms of matrix-valued functions and therefore invariant under
arbitrary unitary transformation of input matrix.  As a result, the
continuation of off-diagonal elements becomes straightforward.
Without introducing any further ambiguity, the Bayesian probabilistic
interpretation is maintained just as in the conventional
maximum entropy method. The applications of our generalized formalism 
to a model spectrum and a real material demonstrate its usefulness and 
superiority.
\end{abstract}

\author{Jae-Hoon Sim}
\affiliation{Department of Physics, Korea Advanced Institute of Science and Technology
(KAIST), Daejeon 305-701, Korea}

\author{Myung Joon Han}
\affiliation{Department of Physics, Korea Advanced Institute of Science and Technology
(KAIST), Daejeon 305-701, Korea}
\email{mj.han@kaist.ac.kr}

\date{\today}
\maketitle

\section{Introduction}

Imaginary time Green's function method such as Quantum Monte Carlo
(QMC) is a main workhorse for various many-body problems
\cite{Gull_RMP_2001, RevModPhys.73.33, PhysRevLett.56.2521,
  PhysRevB.46.560}.  While it has been successful for both impurity
and periodic systems, the output data on the imaginary axis
({\it e.g.}, Matsubara Green's function $G(i\omega_{n})$) should be
transformed to the real-frequency spectrum $A(\omega)$ in order to be
compared with experimental results. Namely, physical observables can only be
accessed indirectly via analytic continuation. While the calculation
of $G(i\omega_{n})$ from $A(\omega)$ is straightforward, its inverse
is an ill-posed problem due to the large conditional number of the
kernel matrix. The small noise in $G(i\omega_{n})$ can lead to large
fluctuations in $A(\omega)$, and the double precision is far from
being enough \cite{PhysRevB.93.075104}. Among well-established
methods, such as Pade
\cite{Vidberg_JLTP_1977,Gunnarsson_PRB_2010_82_165125}, stochastic
method \cite{Sandvik_PhysRevB.57.10287} and others
\cite{Otsuki_PRE_95_061302,PhysRevB.89.245114,Arsenault2017,Haule_PRB_2010_81_195107},
maximum entropy method (MEM) is one of the most widely used
\cite{Bergeron_PRE_94_023303,Jarrell1996,Gunnarsson2010}.

One obvious limitation of conventional MEM is about
the non-diagonal components of Matsubara functions.
Since the conventional formalism
is rigorous only for non-negative and additive functions \cite{LAUE_JMR_1895_63.418}, it has been a
challenge to make the continuation of off-diagonal matrix
elements which can be negative or complex. This limitation
becomes particularly serious when one tries to understand 
the real materials based on, for example, dynamical mean-field theory (DMFT) \cite{Georges_PRB_1992,Zhang_PRL_1993} combined with QMC 
impurity solver \cite{Gull2011,Haule2007_PRB.75.155113}. No matter how correctly the Matsubara
function or self-energy be computed, severely limited is
to understand the electronic property, especially the effect of 
spin-orbit coupling (SOC), crystal-field effect or any other
factors that can generate non-diagonal parts of the functions
\cite{Pavarini2004,Martins_PRL_2011}.

One possible way to overcome this limitation is to transform 
the imaginary frequency data to a `good basis
set' on which the Green's function can be represented as diagonal as
possible and the off-diagonal elements be neglected. Another 
approach tries to relax the non-negativity conditions or 
to construct the auxiliary functions with positive definite property
\cite{Reymbaut_PRB_2015_92_060509,Tomczak_JPCM_19.365206_2007,PhysRevB.96.024438}.
Recently, a notable idea
has been suggested \cite{Kraberger_PRB_2017}. In order to apply MEM to
the off-diagonal elements of spectral function, $A^{{\rm off}}(\omega)$,
Kraberger {\it et al.} decomposed $A^{\rm off}(\omega)$ so that the positive
definite condition be satisfied. It is noted however that 
$SU(N)$ invariance is still not preserved  
in the sense that the resulting spectrum is 
dependent on the basis choice.

In the current study, we generalize MEM by reformulating it
with quantum relative entropy. It can be regarded as a 
quantum version of MEM. Within this maximum quantum
entropy method (MQEM), the matrix is not decomposed nor treated
element-wise, but is directly continued as a single object.
Thus our formalism is guaranteed to be basis-independent.  This
outstanding feature enables us to perform the analytic continuation 
for the off-diagonal parts. We apply MQEM to a
model spectrum, whose ideal spectrum can be known by construction, 
and to a realistic material example of Sr$_{2}$IrO$_{4}$,
for which the full matrix information is essential due to strong
SOC and structural distortion. The results demonstrate 
the usefulness and superiority of our new formulation.

\section{Formalism}

\subsection{Quantum entropy}
Matsubara frequency Green's function $G(i\omega_{n})$ (or self-energy
$\Sigma(i\omega_{n})$) is analytically continued to real-frequency
$G(\omega)$ (or $\Sigma(\omega)$). For a given $G(i\omega_{n})$,
spectral function $A(\omega)=-\frac{1}{\pi}\Im G(\omega+i0^{+})$ is
obtained by inverting the integral equation
\begin{align}
{G}(i\omega_{n}) & =\int d\omega\frac{{A}(\omega)}{i\omega_{n}-\omega}\label{eq:TransFrom}\\
 & =\int d\omega{\bf K}(i\omega_{n},\omega){A}(\omega).
\end{align}
Note that both ${G}(i\omega_{n})$ and ${A}$ are in general
matrix-valued functions. A kernel $\boldsymbol{K}(i\omega_{n},\omega)$
is ill-conditioned and the direct inverse of $A={\bf K}^{-1}G$ or
the minimization of 
\begin{align}
\chi^{2} & =\frac{1}{2}\sum_{n}\norm{{G}(i\omega_{n})-\int d\omega{\bf K}(i\omega_{n},\omega){A}(\omega)}^{2}\label{eq:chi2}
\end{align}
is not quite feasible, likely leading to the violation of
non-negativity condition ($A_{ii}>0$) and sum rule ($\int d\omega
A_{ii}=1$).

To solve this ill-posed problem, MEM introduces entropy $S$ for
diagonal components,
\begin{equation}\label{class_S}
S[A(\omega)||D(\omega)]=\int d\omega A(\omega)\ln\frac{A(\omega)}{D(\omega)},
\end{equation}
which is also known as Kullback-Leibler distance
\cite{Bergeron_PRE_94_023303,Jarrell_book}.  $D(\omega)$ is a default
model providing the essential features of spectra, which can be
determined by `annealing' procedure
\cite{Bergeron_PRE_94_023303,Jarrell_book} or by making use of high-frequency behavior of input data \cite{Bergeron_PRE_94_023303}. In
MEM, it is the `free energy' $F=\chi^{2}+\alpha S$ (not $\chi^{2}$
in Eq.~(\ref{eq:chi2})) that is minimized with a fitting parameter
$\alpha$, and the entropy $S$ requires the positiveness of the spectrum
\cite{Kraberger_PRB_2017}.

Hereafter, we use a hat~(~$\hat{ }$~) notation to emphasize the matrix
values. In order to consider the whole matrix continuation (not
element-decomposed) and not to lose any off-diagonal information, we
first renormalize Matsubara function by
$\hat{G}\to\hat{G}/\Tr[\hat{M_{0}}]$.  Here $M_0 = \int d\omega
\hat{A}(\omega)$ is zeroth moments of the spectrum.  With renormalized
Matsubara function, the asymtotic behavior can be written as
$\hat{G}(i\omega_{n}\to\infty)=\hat{M_{0}}/i\omega_{n}$ with
$\Tr\hat{M_{0}}=1$.  Second, we divide $\hat{A}(\omega)$ into two
parts, $\hat{A}(\omega)=P(\omega)\hat{\rho}(\omega)$ and
$P(\omega)=\Tr\hat{A}(\omega)$.  Here the sum rule is written as
\begin{equation}
\int d\omega\Tr\hat{A}=\int d\omega P(\omega)=\Tr\text{\ensuremath{\hat{M}_{0}}}=1.
\label{eq:sum_rule}
\end{equation}
Thus, within this formalism, $P(\omega)$ can be interpreted
as a classical probability distribution and $\hat{\rho}(\omega)$
a density matrix.

Now we extend the entropy of Eq.~(\ref{class_S}) to quantum relative
entropy, which is widely used in the non-equilibrium thermodynamics
\cite{Kawai_PRL_98_080602_2007,KHKim_PRE_2011_84_012101} as well as
the information science \cite{Nielsen2010}:
\begin{align}
S^{Q}(\hat{A}||\hat{D}) & =\int d\omega\Tr\hat{A}(\omega)(\ln\hat{A}(\omega)-\ln\hat{D}(\omega))\\
 & =S[P(\omega)||D(\omega)]+\int d\omega P(\omega)S^{Q}[\hat{\rho}(\omega)||\hat{\sigma}(\omega)].\nonumber 
\end{align}
Here the default model $\hat{D}(\omega)=D(\omega)\hat{\sigma}(\omega)$
is further decomposed into $D(\omega)=\Tr\hat{D}(\omega)$. While the
first term in the second line, $S[P(\omega)||D(\omega)]$, is the
classical entropy used in MEM (see Eq.~(\ref{class_S})), the second
term is introduced to regularize matrix elements. In our
formalism, the free energy functional to be minimized is defined by
the matrix-valued functions; $F[\hat{A};\hat{G}]=\chi^{2}-\alpha
S^Q(\hat{A}||\hat{D})$.  We stress that this free energy functional
is $SU(N)$ invariant, i.e.,
\begin{equation}
F[U\hat{A}U^{\dagger};U\hat{G}U^{\dagger}]=F[\hat{A};\hat{G}]
\end{equation}
for unitary matrix $U$.

While our formalism assumes that the spectrum
$\hat{A}$ is Hermitian, any non-Hermitian spectrum $\hat{A}_{\rm
  nonH}$ can be divided into two Hermitian matrices;
\begin{align}
\hat{A}_{\rm nonH}^{R} & =(\hat{A}_{\rm nonH}+\hat{A}_{\rm nonH}^{\dagger})/2\\
\hat{A}_{\rm nonH}^{I} & =(\hat{A}_{\rm nonH}-\hat{A}_{\rm nonH}^{\dagger})/2i.
\end{align}
And therefore, $\hat{A}_{\rm nonH}^{R}$ and 
$\hat{A}_{\rm nonH}^{I}$ can be dealt with separately.

\subsection{Iterative equation}

The key task is to minimize free energy $F$:
\begin{align}
\Phi & =\min_{A(\omega)}F[\hat{A}(\omega)]\\
 & =\min_{A(\omega)}[\frac{1}{2}\chi^{2}+\alpha S^Q(\hat{A}||\hat{D})].\nonumber 
\end{align}
This minimization can be conducted by using the stationary condition,
${\delta F}/{\delta\hat{A}}=0$. With a trace norm
$\norm*{\hat{M}}=\Tr(\hat{M}^{\dagger}\hat{M})$ for
Eq.~(\ref{eq:chi2}), a set of self-consistent equations is given as
follows:
\begin{equation}
\hat{H}_{\omega}\ket{\psi_{\omega}}=\epsilon_{\omega}\ket{\psi_{\omega}}\label{eq:SC_equation},
\end{equation}
where 
\begin{widetext}
\[
\hat{H}_{\omega}[\hat{A}(\omega)] =
\frac{1}{2}\sum_{i\omega}K^{*}(i\omega_{n},\omega)\left(\hat{G}(i\omega_{n})
- \int d\omega^{\prime}K(i\omega_{n},\omega^{\prime})\hat{A}(\omega^{\prime})\right)
+ \frac{\alpha}{2}\ln\hat{D}(\omega)+{\rm {h.c.}},
\]
\end{widetext}
and 
\begin{equation}
\hat{A}(\omega)=\int d\epsilon\frac{1}{Z}e^{-\epsilon_{w}/ \alpha}\ketbra{\psi_{\omega}}{\psi_{\omega}}.
\end{equation}
Then Eq.~(\ref{eq:SC_equation}) can be solved iteratively. Note that
these equations represent a quantum system described by Hamiltonian
$\hat{H}_{\omega}[\hat{A}(\omega)],$ and its spectrum is given by
density matrix of the canonical ensemble with temperature $\alpha$. It
is not surprising since MEM can be regarded as a mean-field
realization of stochastic approximation (SA) \cite{Beach2004}.

We used Pulay mixing scheme \cite{PULAY1980393} and its generalization
\cite{BANERJEE201631} to achieve the stable convergence.  The results were
compared to the solution of the reduced independent variables in the
singular space \cite{Kraberger_PRB_2017,Bryan_EurBioPhy_19990}. To
minimize the real-frequency grid size, cubic splines in combination
with non-uniform real-frequency grids have been adopted
\cite{Bergeron_PRE_94_023303}. For more details, see
Appendix~\ref{app-A}.

\subsection{Default model}

We take Gaussian shape of default model to avoid the data noise.  The
asymptotic behavior of high-frequency data determines the first a few
moments of spectra \cite{Bergeron_PRE_94_023303}:
\begin{equation}
  \hat{G}(i\omega_{n}) =
  \frac{\hat{M}_{0}}{i\omega_{n}} +
  \frac{\hat{M}_{1}}{(i\omega_{n})^{2}} +
  \frac{\hat{M}_{2}}{(i\omega_{n})^{3}} + ...,
\end{equation}
where $\hat{M}_{j}=\int\omega^{j}\hat{A}(\omega)d\omega$ is the $j$-th
moment of $\hat{A}(\omega)$. To define Gaussian curves for given
moments $M_{j}$ ($j=0,1,2$) is straightforward in the scalar version
of MEM. In the matrix formalism of our MQEM, on the other hand,
finding out the analytic solution is not quite feasible due to the
fact that $\hat{M}_{j}$ does not commute in general with each other;
$[M_{j},M_{j\prime}]\neq0$.

Here we propose a way to find out the `featureless' default models 
for a few given moments, $M_{j}$ ($j=0,1,2$). 
Recalling that Gaussian curve has
the maximum entropy among the distributions with a specified variance,
we define a default model that maximizes
\begin{equation}
  S_{D} = \int d\omega\hat{D}(\omega)\ln\hat{D}(\omega) +
  \sum_{j=0}^{2}\Tr\hat{\mu}_{j}\int\omega^{j}\hat{D}(\omega)d\omega.
\end{equation}
where $\hat{\mu}_{j}$ is Lagrange multiplier introduced by the
constraint $\int\omega^{j}\hat{D}(\omega)d\omega=M_{j}$.
The stationarity condition 
$\partial S_{D}/\partial\hat{D}(\omega)=0$ reads
\begin{equation}
\hat{D}(\omega) = \exp(\sum_{j=0}^{2}\hat{\mu}_{j}\omega^{j}).
\end{equation}

\subsection{Fitting parameter $\alpha$}

\begin{figure}
\includegraphics[width=1\columnwidth]{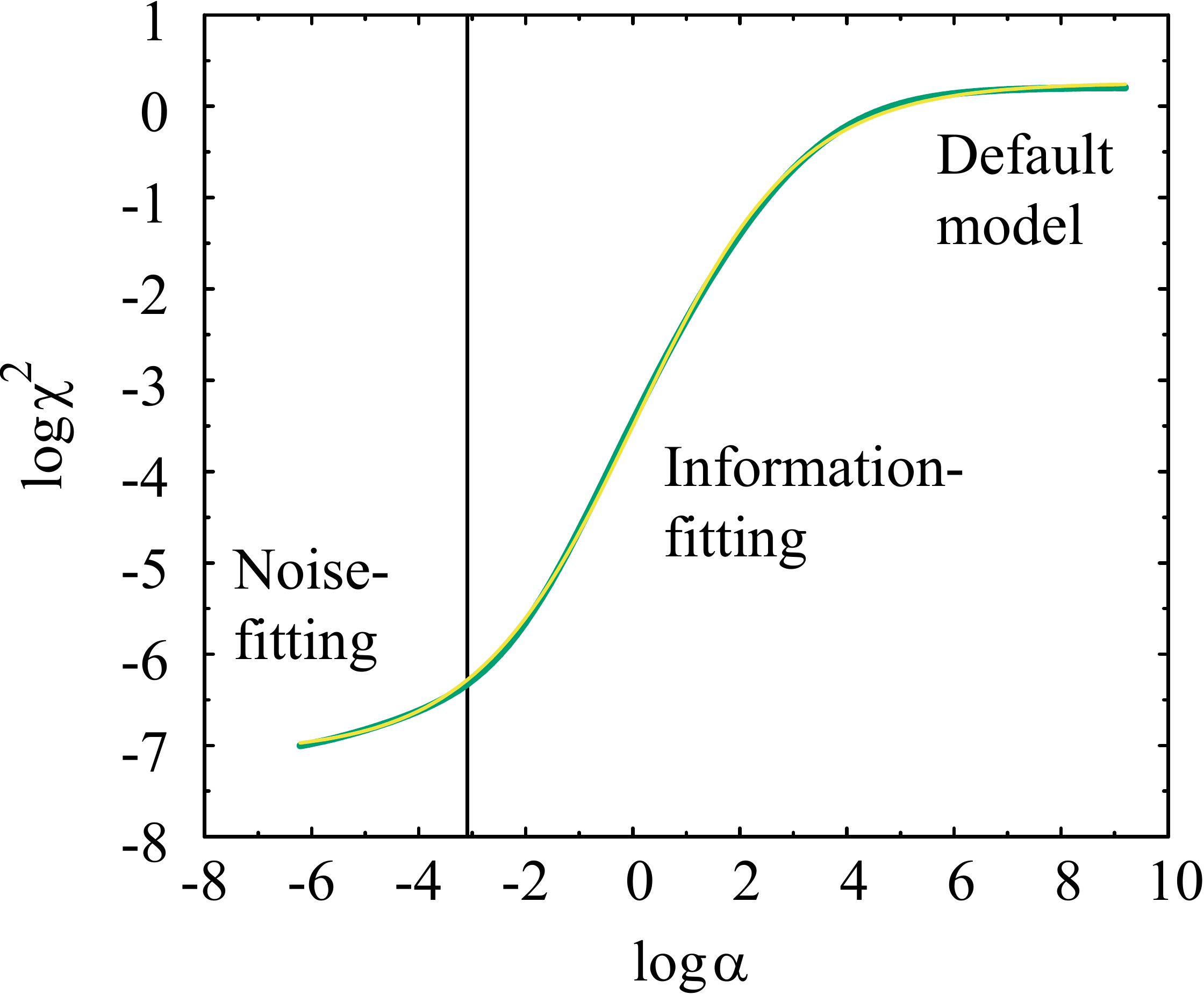}
\caption{The green line presents log~$\chi^{2}$ 
	as a function of $\log\alpha$ in the case
  of model spectrum discussed in Sec.\ref{subsec:Model_spectrum}.
  Gaussian noise of $\sigma=10^{-3}$ is
  introduced to each element of Green's function matrix.
  Yellow line is the fitting curve with
  optimal $\log\alpha_{{\rm opt}}=-3.09$ (black vertical line).\label{fig:optimal_alpha}}
\end{figure}

A popular approach to calculate spectral functions is to optimize the
parameter $\alpha$ by a statistical method within the probabilistic
interpretation of MEM \cite{Jarrell1996}.  Alternatively,
an average value of the spectra calculated by many different $\alpha$
values  can be taken
\cite{Bergeron_PRE_94_023303,Kraberger_PRB_2017}. Recently, a
different approach has been suggested \cite{Bergeron_PRE_94_023303}.
In this approach $\log\chi^{2}$ is computed 
as a function of $\log\alpha$, and
two different regions (namely, `information-fitting' and
`noise-fitting' region) are considered. 
The optimal $\alpha$ is then
determined at the maximum curvature of
$\log\chi^{2}(\log(\alpha))$. We used a similar approach 
in our MQEM implementation.  We fit
$\log\chi^{2}(\log(\alpha))$ curve by Fermi-Dirac function as shown in
Fig.~\ref{fig:optimal_alpha}. The optimal $\alpha$ is determined
by the maximum second deviation of the fitting function. A clear
advantage of this technique is the numerical stability against the
grid changes.

\section{Result and Discussion}

\subsection{Simple model spectrum \label{subsec:Model_spectrum}}

As the first example, we apply our method to a simple model system.
The Green's function $\hat{G}^{{\rm in}}(i\omega_{n})$ is
obtained from a model spectral function which is given by a $2\times2$
matrix:
\begin{equation}
\hat{A}(\omega)=\hat{R}\left[\begin{pmatrix}1 & 0\\
0 & 0
  \end{pmatrix} ( e^{-\frac{1}{2}(\frac{\omega+\omega_{0}}{2})^{2}}
  + e^{-\frac{1}{2}(\frac{\omega-\omega_{0}}{2})^{2}})\right]
\hat{R}^{\dagger}.
\label{eq:model_spectrum}
\end{equation}
Note that obtaining $\hat{G}^{{\rm in}}(i\omega_{n})$ from
$\hat{A}(\omega)$ is not ill-conditioned.  Here the two-peak Gaussian
spectrum centered at $\omega_{0}=\pm1.5$ is rotated by a rotation
matrix $\hat{R}=\begin{pmatrix}\cos(\theta) &
i\sin(\theta)\\ i\sin(\theta) & \cos(\theta)
\end{pmatrix}$ with $\theta=(2\pi\omega/T_{\omega})^{2}$. For $T_{\omega}=\infty$
($\theta=0),$ the spectral function corresponds to the trivial case
that off-diagonal elements are all zero. At finite $T_{\omega}$,
$\hat{A}(\omega)$ has non-zero off-diagonal values.  In performing MEM
continuation, we also introduced random Gaussian noises to the Green's
functions with a standard deviation of $\sigma=10^{-4}$ in order to
mimic a realistic QMC situation.

\begin{figure}
\includegraphics[width=1\columnwidth]{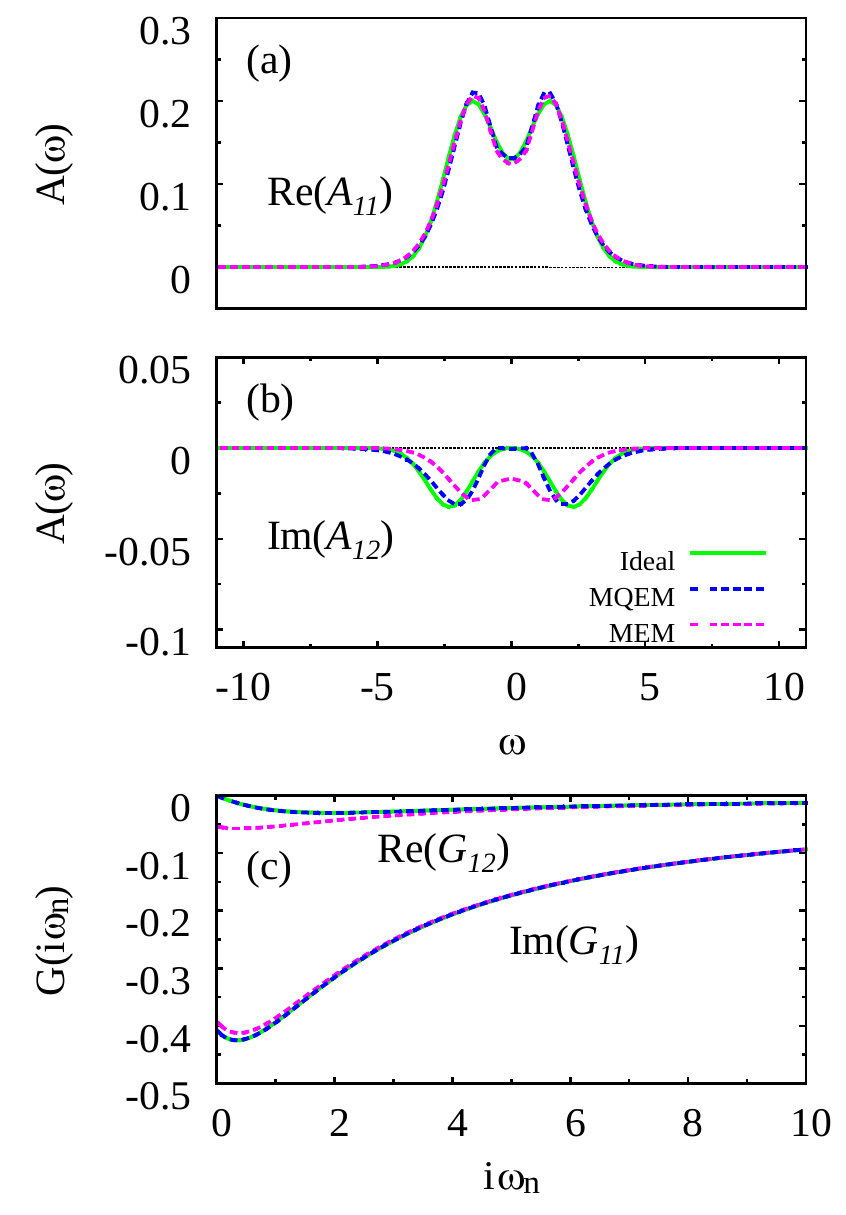}
\caption{(a, b) The calculated spectral function $\hat{A}(\omega)$ by
  using the conventional MEM (magenta) and our method (blue). The
  ideal spectra, from which $G(i\omega_n)$ is calculated, are also
  presented for comparison (green).  The diagonal and off-diagonal
  components are presented in (a) and (b), respectively.
  The off-diagonal components are dominated by imaginary part 
  due to the form of the rotation matrix $\hat{R}$.
  (c) The input Green's
    function in Matsubara frequency axis (green) and the Green's
    function reconstructed from $\hat{A}(\omega)$ using
    Eq.~(\ref{eq:TransFrom}) (blue and magenta). The solid 
  (below) and dashed lines (above) correspond to $\Im[G_{11}(i\omega_{n})]$ and $\Re[G_{12}(i\omega_{n})]$, respectively. \label{fig:gMEM_model}}
\end{figure}

Figure~\ref{fig:gMEM_model} shows the calculated spectra from
the input of Eq.~(\ref{eq:model_spectrum}).
The conventional MEM and the generalized MQEM results are presented in
magenta and blue lines, respectively, along with the ideal spectrum (green)
from which the input Green's function $\hat{G}^{{\rm in}}(i\omega_{n})$
is generated.  It is noted that the conventional MEM does not
well reproduce the off-diagonal part of spectral function
(Fig.~\ref{fig:gMEM_model}(b)) while the diagonal part is in good
agreement with the ideal spectrum (Fig.~\ref{fig:gMEM_model}(a)).
This is a well-known limitation of MEM. Here the results of
conventional MEM are obtained from the properly-chosen basis set in
which the off-diagonal components of $\hat{G}^{{\rm in}}(i\omega_{n})$ are
minimized; {\it i.e.,} min~$\sum_{i\omega_{n}}\sum_{i\neq
  j}\left|\hat{G}_{ij}^{{\rm in}}(i\omega_{n})\right|^{2}$ 
($\theta=0.972$ rad). Note that,
even with this `best' basis, the off-diagonal elements are significantly
deviated from the ideal result as shown in
Fig. \ref{fig:gMEM_model}(b).
The same feature is also observed in
$G(i\omega_{n})$, see Fig. \ref{fig:gMEM_model}(c).
The conventional MEM result shows the noticeable deviation from the 
ideal (or original) curve especially for the off-diagonal part. Note that,
in this example, there is no unitary transformation for the basis
set on which the matrix-valued $\hat{A}(\omega)$ (or equivalently
$\hat{G}^{{\rm in}}(i\omega_{n})$) is diagonalized at all frequencies,
and therefore the conventional MEM has no way to be satisfactory.

A remarkable improvement is clearly noticed in our result of MQEM.
Even for the off-diagonal components, the generalized MQEM results are
in good agreement with the ideal spectrum; see
Fig.~\ref{fig:gMEM_model}(a) and (b). The excellent agreement is also
found for $G(i\omega_{n})$ as shown in Fig.~\ref{fig:gMEM_model}(c).
This result of simple model spectrum demonstrates the capability of
MQEM for the continuation of matrix-valued functions.

\subsection{Real material example: Sr$_{2}$IrO$_{4}$}

\begin{figure}
\includegraphics[width=0.9\columnwidth]{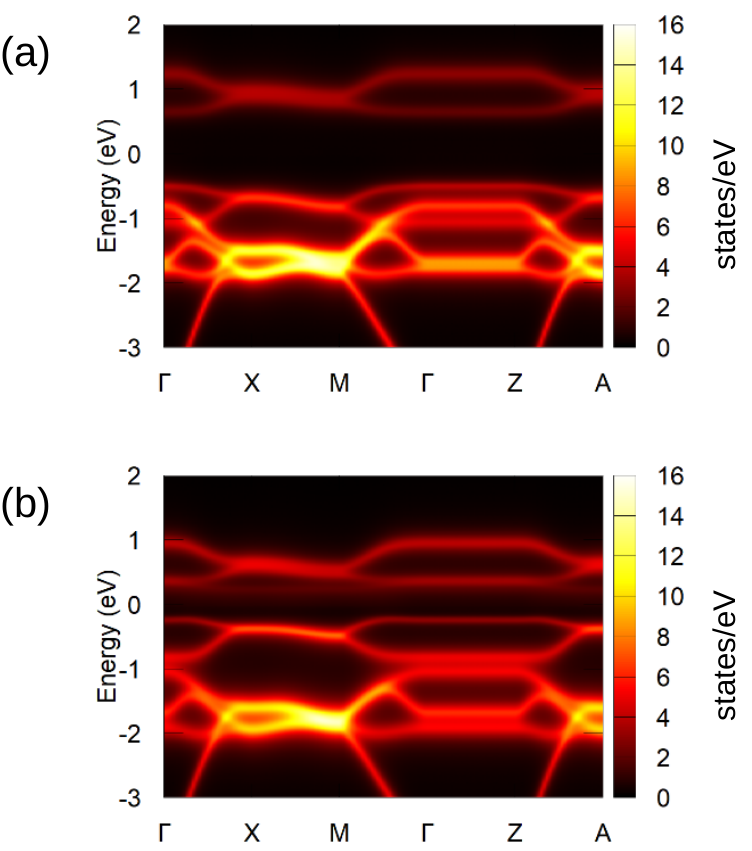}\caption{The calculated
  DMFT spectral function of Ir-$t_{2g}$ states obtained by (a) our
  generalized MEM and (b) the conventional MEM. The chemical potential
  is set to be zero energy.
\label{fig:dmft_band_SIO214}}
\end{figure}

As a real material example, we consider Sr$_{2}$IrO$_{4}$. The local
Green's function and self-energy of this material are featured by the
significant off-diagonal components caused by strong SOC and
structural distortions.  Thus, dealing properly the off-diagonal
elements is of crucial importance to describe its electronic
structure. We calculate Matsubara functions by LDA+DMFT (local density
approximation plus dynamical mean-field theory) method based on
Wannier-projected $t_{2g}$ orbitals \cite{Mazari1997,Souza2001}. The interaction
parameter of $U=2.2$ eV is adopted \cite{Martins_PRL_2011}.  Further
computation details can be found in Appendix~\ref{app-B}.  Analytic
continuation of impurity self-energy $\Sigma(i\omega_{n})$ is
conducted to obtain $\Im\Sigma(\omega)$, and the real part is obtained
by Kramers-Kronig transformation.

Real frequency self-energy can be obtained via the analytic
continuation of Weiss field, ${\mathcal{G}}_{0}(i\omega_{n})$, and
impurity Green's function, $G_{{\rm imp}}(i\omega_{n})$.  Self-energy
on the real-frequency axis is then given by Dyson's equation,
$\Sigma(\omega) = {\mathcal{G}}^{-1}_{0}(\omega) - G^{-1}_{{\rm
    imp}}(\omega)$ \cite{Wang_PRB_2009}.  In practice, 
widely used is to perform the continuation of auxiliary Green's
functions which are constructed from the self-energy
\cite{Martins_PRL_2011,Mravlje_PRL_2016,Kraberger_PRB_2017,
  Zhang2013}.  While there are many different ways to construct the
auxiliary Green's functions, we perform the continuation of
$\hat{\Sigma}^{{\rm
    dyn}}(i\omega_{n})=\hat{\Sigma}(i\omega_{n})-\hat{\Sigma}(i\infty)$
\cite{Wang_PRB_2009}.  It is noted that the element-wise MEM is not
quite feasible for $\hat{\Sigma}^{{\rm dyn}}$ due to the fact that the
high-frequency behavior of the off-diagonal components of
$\hat{\Sigma}^{\rm dyn}$ is proportional to ${1}/{i\omega_n}$ with the
finite norm of the spectral function \cite{Kraberger_PRB_2017}.  We
emphasize that our formalism is free from this deficiency and provides
the full matrix information of high-frequency coefficents; see
Eq.~(\ref{eq:sum_rule}).

The result of MQEM is presented in Fig.~\ref{fig:dmft_band_SIO214}(a).
The calculated spectral function $A({\bf k},\omega)$ is in reasonable
agreement with the well-known features of this material including the
relative position of so-called $j_{\rm eff}$=1/2 
and $j_{\rm eff}$=3/2 bands
\cite{Wang2013_PRB.87.245109,PhysRevLett.101.226402,Kim2008,PhysRevLett.101.226402}.

MQEM result is significantly different
from that of conventional MEM. By comparing
Fig.~\ref{fig:dmft_band_SIO214}(a) and (b), the differences are
clearly noticed.  For example, the separation between the conduction
and valence band states is markedly enhanced in MQEM
(Fig.~\ref{fig:dmft_band_SIO214}(a)) and therefore the band gap 
becomes larger.
While $j_{{\rm eff}}$=1/2 states (upper and lower Hubbard band) 
moves away from Fermi level, the
$j_{{\rm eff}}$=3/2 states do not show a significant change. 
It is likely due to that 
$\Sigma^{\rm dyn}$ is dominated by the static Hartree terms
in this fully occupied states.

Our results show that taking the full account of off-diagonal matrix
elements is important to correctly describe the electronic
structure. While the effect of ignoring off-diagonal part can be
minimized by taking better basis set rather than
Ir-$t_{2g}$\cite{Martins_PRL_2011,Zhang2013}, it is not always
straightforward to make the right choice. In many different
situations and due to many different reasons, the off-diagonal elements can become non-negligible. Therefore it is important to take all matrix information
through the continuation process.

It can be an interesting future direction to further extend the idea of MQEM.
Introducing quantum entropy can extend the physical implication and the applicability
of currently available methods or techniques, 
especially in our case for dealing with the off-diagonal information.
The similar idea might be applicable to the 
other non-Hermitian matrix-valued functions such as Gorkov's Green's function 
for superconducting order parameter \cite{Reymbaut_PRB_2015_92_060509}.

\section{Summary}

By introducing quantum relative entropy we resolve a 
long-standing issue of analytic continuation, namely,
the continuation of off-diagonal matrix elements.
Based on quantum relative entropy, the 
functions are treated as being matrix-valued and 
the non-negativity condition as well as the sum rule
are extended.
The invariance under unitary transformation
and the Hermiticity of spectral function $\hat{A}(\omega)$
are inherently satisfied in the general context.
As a result, it becomes straightforward to perform
analytic continuation of the off-diagonal as well as diagonal
components without any further approximation or ad-hoc
treatment. The capability and usefulness 
of our method is demonstrated
by two examples. In both of model spectrum and
a real material example of Sr$_2$IrO$_4$, our MQEM
provides a reliable description of off-diagonal elements
which cannot be well treated within the conventional schemes.

\begin{acknowledgments}
We thank Junya Otsuki, Hongkee Yoon and Hunpyo Lee 
for useful comment and discussion. This work was supported by
Basic Science Research Program through the National Research Foundation
of Korea (NRF) funded by the Ministry of Education 
(2018R1A2B2005204).
\end{acknowledgments}

\appendix

\section{Implementation details}
\label{app-A}

\subsection*{Kernel matrix for cubic spline spectral function}

For a given discrete grid set of frequencies $\omega_{j}\in[-W,W]$,
Eq.~(\ref{eq:TransFrom}) reads
\begin{equation}
  G(i\omega_{n}) \approx
  \sum_{j=1}^{N_{\omega}+1} \frac{A(\omega_{j})}{i\omega_{n}-\omega_{j}}
  \Delta\omega_{j}.
\end{equation}
In order to achieve a high accuracy with a decent number of grids,
we adopted the cubic spline interpolation. The coefficients of cubic polynomials,
$S_{j}(\omega)=a_{j}(\omega-\omega_{j})^{3}+b_{j}(\omega-\omega_{j})^{2}+c_{j}(\omega-\omega_{j})+d_{j}$
(for $\omega_{j}<\omega<\omega_{j+1}$), are the
solution of the linear equations:
\begin{widetext}
\begin{eqnarray}
S_{j}(\omega_{j}) & = & d_{j}=A(\omega_{j})\\
S_{j}(\omega_{j+1}) & = & a_{j}\Delta\omega_{j}^{3} +
b_{j}\Delta\omega_{j}^{2} + c_{j}\Delta\omega_{j}+d_{j} =
A(\omega_{j+1})\\
S_{j}^{\prime}(\omega_{j+1}) - S_{j+1}^{\prime}(\omega_{j+1})
& = &
3a_{j}\Delta\omega_{j}^{2}+2b_{j}\Delta\omega_{j}+c_{j}-c_{j+1}
=0\\
S_{j}^{\prime\prime}(\omega_{j+1}) - S_{j+1}^{\prime\prime}(\omega_{j+1})
& = &
6a_{j}\Delta\omega_{j}+2b_{j}-2b_{j}=0.
\end{eqnarray}
\end{widetext}
Here $j=1,..., N_{\omega}$ in (A.2) and (A.3), and
$j=1,...,N_{\omega}-1$ in (A.4) and (A.5), providing ($4N_{\omega}-2$)
equations. Two more equations are from boundary conditions:
\begin{equation}
S_{1}^{\prime\prime}(\omega_{1}) =
S_{N_{\omega}}^{\prime\prime}(\omega_{N_{\omega}+1}) = 0.
\end{equation}
Thus the transformation matrix $T$ is obtained and it gives 
rise to a vector $\Gamma=TA$ for the spline coefficients
in terms of the spectral function $A$ at the grid points
\cite{Bergeron_PRE_94_023303};
$A_i = A(\omega_i)$ and $\Gamma = (a_1, b_1, c_1, d_1, ..., d_{N_\omega})$.

With the known coefficients, Eq.~(\ref{eq:TransFrom}) can be 
rewritten as
\begin{align}
G(i\omega_{n}) & = \sum_{j=1}^{N_{\omega}} \int_{\omega_{j}}^{\omega_{j+1}} d\omega
\frac{S_{j}(\omega)}{i\omega_{n}-\omega}\label{eq:KernelCubic}\\
& =K\Gamma.
\end{align}
Here $K$ is the matrix obtained by integrating
Eq.~(\ref{eq:KernelCubic}) \cite{Bergeron_PRE_94_023303}.
Finally, we have
\begin{equation}
G=\boldsymbol{K}A=KTA.
\end{equation}
This procedure provides us the more stable and efficient numerics
for MQEM.

\subsection*{Non-uniform real-frequency gird}

To reduce the number of grid points, non-uniform real-frequency
grid technique is adopted \cite{Bergeron_PRE_94_023303} in which
three different regions are considered as the grid sections;
$W_{L}=[\omega_{{\rm min}}, w_{l})$, 
$W_{C}=[w_{l}, \omega_{r}]$, and 
$W_{R}=(\omega_{r}, \omega_{{\rm max}}]$. For
the central region $W_{C}$, we take a
regular grid spacing of $\Delta\omega$. For
$W_{L}$, on the other hand, the grid is defined by
\begin{equation}
\omega_{j}=\frac{1}{u_{j}}+\omega_{0l}\in W_{L}
\end{equation}
where $j=1,..., N_{L}$. The free parameters $N_{L}$, $u_{j}$, and
$\omega_{0l}$ are to be determined. By assuming a constant step $\Delta
u$, $\omega_{l}=\frac{1}{u_{N_{L}+1}}+\omega_{0l}$ and
$\omega_{N_{L}}=\omega_{l}-\Delta\omega$, we have
\begin{align}
\Delta u & =u_{N_{L+1}}-u_{N_{L}}\\
 & =\frac{1}{\omega_{l}-\omega_{0l}}-\frac{1}{\omega_{l}-\Delta\omega-\omega_{0l}},
\end{align}
$u_{j}=j\Delta u$, and
$N_{L}\Delta u=\frac{1}{\omega_{l}-\Delta\omega-\omega_{0l}}$.
With a given $\Delta u$,
\begin{align}
\omega_{{\rm min}} & =\frac{1}{\Delta u}+\omega_{0l}\\
& = -\frac{(\omega_{l}-\Delta\omega-\omega_{0l})
  (\omega_{l}-\omega_{0l})}{\Delta\omega}
+ \omega_{0l},
\end{align}
and 
$\omega_{0l}=\omega_{l}+\sqrt{\Delta\omega(\omega_{l}-\omega_{{\rm
      min}})}$.  Since we have an integer value of
\begin{equation}
  N_{L} = {\rm ceil}
  (\frac{1}{(\omega_{l}-\Delta\omega-\omega_{0l})\Delta u})
  = {\rm ceil}(\frac{\omega_{0l}-\omega_{l}}{\Delta\omega}),
\end{equation}
we re-define $\omega_{0l}=\omega_{l}+N_{L}\Delta\omega_{l}$ and
$\omega_{{\rm min}}$. The same numerical approach is also used 
for $W_{R}$.

\section{Calculational details of LDA+DMFT}
\label{app-B}

First-principles electronic structure calculations have been carried
out based on DFT (density functional theory) within LDA (local density
approximation) \cite{Ceperley_PRL_1980}. We used our DFT software package
`OpenMX' \cite{Han2006,Ozaki2004,Ozaki2003,openmx} for Sr$_2$IrO$_4$.
8$\times$8$\times$1 $\vb{k}$-points for the slab geometry have been taken.
SOC is treated within a fully relativistic $j$-dependent formalism
\cite{Macdonald_JPCSSP_1979}. To describe the electronic correlation,
single-site DMFT has been adopted
\cite{Georges_PRB_1992,Zhang_PRL_1993}.  The correlated subspace was
constructed by maximally localized Wannier functions starting
from the initial projections onto the atomic Ir-$t_{2g}$ orbitals
\cite{Mazari1997,Souza2001}.  This Hamiltonian serves as the
non-interacting $H_{0}$ for the multi-band Hubbard Hamiltonian
$H=H_{0}+H_{{\rm int}}$. The interaction part is expressed in the
Slater-Kanamori form of $H_{{\rm int}}=\sum_{i}h_{i,{\rm int}}$;
\begin{equation}
  h_{i,{\rm int}} = \sum_{\alpha} Un_{i\alpha\uparrow}n_{i\alpha\downarrow}
  +\sum_{\alpha\neq\beta}U^{\prime}n_{i\alpha\uparrow}n_{i\beta\downarrow}
\end{equation}
where $U$ and $U^{\prime}$ refers to the intra-orbital and inter-orbital
interaction, respectively, and $U^{\prime}=U=2.2$ eV. 
Hund interaction $J_{H}$ is set to zero which does not change
any of our main conclusions. The
Hamiltonian is solved within single-site DMFT (dynamical mean-field
theory) by employing a hybridization expansion continuous-time quantum
Monte Carlo (CT-QMC) \cite{Haule2007_PRB.75.155113} with
$2.24\times10^{8}$ measurements. In this procedure, local Green's
functions are calculated using momentum-independent self-energy;
\begin{equation}
  G_{{\rm loc}}(i\omega_{n}) = \frac{1}{N_{k}}
  \sum_{{\bf k}}\frac{1}{i\omega_{n}+\mu-H_{0}({\bf k})-\Sigma(i\omega_{n})},
\end{equation}
where $H_{0}({\bf k})$ and $\Sigma(i\omega_{n})$ are given by
$12\times12$ matrices. Self-energy is decomposed into $6\times6$
matrices corresponding to two Ir sites,
$\Sigma(i\omega_{n})=\Sigma_{{\rm
    Ir(1)}}(i\omega_{n})\oplus\Sigma_{{\rm Ir(2)}}(i\omega_{n})$.

\bibliographystyle{apsrev4-1}
\bibliography{mem_ref_bib}


\end{document}